\documentstyle[12pt]{article}
\font\goth=eufm10
\font\double=msym10
\font\ccal=cmsy8

\def\cc{{\hbox{\double C}}}

\def\zz{{\hbox{\double Z}}}
\def\aa{{\cal A}}

\def\dd{{\cal D}}
\def\gg{\hbox{\goth g}}
\def\hh{{\cal H}}
\def\hhh{{\hbox{\double H}}}
\def\ff{{\cal F}}
\def\mm{{\hbox{\ccal M}}}
\def\sss{{\cal S}}
\def\t{\,{\rm tr}\,}

\def\ddd{\,\hbox{$\partial\!\!\!/$}}
\def\dee{\,\hbox{\rm D}}
\def\de{\,\hbox{\rm d}}

\def\lb{\left[}
\def\rb{\right]}

\def\ul{\underline}
\def\ot{\otimes}
\def\op{\oplus}

\def\bb{\begin{eqnarray}}
\def\ee{\end{eqnarray}}
\def\eee{\nonumber\end{eqnarray}}
\def\pp{\pmatrix}
\def\qq{\quad}

\begin{document}

\hsize 17truecm
\vsize 24truecm
\font\twelve=cmbx10 at 13pt
\font\eightrm=cmr8
\baselineskip 18pt

\begin{titlepage}

\centerline{\twelve CENTRE DE PHYSIQUE THEORIQUE}
\centerline{\twelve CNRS - Luminy, Case 907}
\centerline{\twelve 13288 Marseille Cedex}
\vskip 4truecm

\centerline{\twelve The Standard Model
\`a la Connes-Lott}

\bigskip

\begin{center}
{\bf Daniel KASTLER}
\footnote{ and Universit\'e d'Aix-Marseille II} \\
\bf Thomas SCH\"UCKER
\footnote{ and Universit\'e de Provence} \\
\end{center}

\vskip 2truecm
\leftskip=1cm
\rightskip=1cm
\centerline{\bf Abstract}

\medskip

The relations among coupling constants
and masses in the standard model \`a la Connes-Lott
with general scalar product are computed in detail.
We find a relation between the top and the Higgs
masses. For $m_t=174\pm22\ GeV$ it yields
$m_H=277\pm40\ GeV$. The Connes-Lott theory
privileges the masses $m_t=160.4\ GeV$ and
$m_H=251.8\ GeV$.

\vskip 2truecm
PACS-92: 11.15 Gauge field theories\\
\indent
MSC-91: 81E13 Yang-Mills and other gauge theories

\vskip 3truecm

\noindent dec. 1994

\noindent CPT-94/P.3091\\
hep-th/9412185

 \end{titlepage}

By now the standard model of electro-weak and strong
interactions in the setting of non-commutative
geometry \cite{cl} is well documented
\cite{cl,k,ks,vg,cG,kGU} and needs no further
introduction. The main virtue of non-commutative
geometry in the context of a Yang-Mills-Higgs model
is that the entire Higgs sector including the choice of
the scalar representation has one common geometrical
origin and is not just added by hand.

The action of a Yang-Mills-Higgs model consists of five
pieces: the Yang-Mills action, the Klein-Gordon action,
the integrated Higgs potential, the Dirac action, and
the Yukawa couplings. Of these only the first has a
genuine geometrical interpretation. As a consequence,
the representation of the gauge potentials is not
arbitrary, it is the adjoint representation. Also, the
cubic and quartic self-couplings of the gauge potentials
are not arbitrary, they are computed from the gauge
invariant scalar product on the Lie algebra and the
structure constants. Similarly, the couplings of the
gauge potentials to the scalars and fermions in the
Klein-Gordon and Dirac actions are fixed by geometry,
`minimal couplings'. All the other coupling constants,
the quadratic, cubic, and quartic ones in the Higgs
potential, and the trilinear Yukawa couplings are
arbitrary except for gauge invariance. Also, the scalar
and the left and right handed fermion representations
are arbitrary.

The action of a Connes-Lott model consists of only two
pieces, the non-commutative Yang-Mills and Dirac
actions. When expanded in terms of ordinary fields the
non-commutative gauge potential consists of the
ordinary gauge potential in the adjoint representation
and a scalar field in a representation, that is computed.
At the same time, the non-commutative Yang-Mills
action yields the ordinary Yang-Mills action, the
Klein-Gordon action and the Higgs potential. Just as
the self-couplings of the gauge potentials, the
self-couplings of the scalars are now computed
from an invariant scalar product in the
non-commutative sense and the underlying algebraic
structure. Finally the non-commutative Dirac action
produces the ordinary Dirac action and the Yukawa
couplings. Input of a Connes-Lott model is a finite
dimensional involutive algebra, the two fermion
representations and their mass matrix. These data then
produce a very particular Yang-Mills-Higgs model
\cite{is2}. Its gauge group is the group of unitary
elements in the algebra or a subgroup thereof. This
model features constraint gauge couplings and a
fixed scalar representation.
Its gauge and scalar boson masses are determined in
terms of the fermion masses. For the standard model,
the scalar representation comes out to be a weak isospin
doublet and with the simplest scalar product one has
\cite{cl34,ks,vg}
\bb g_3=g_2, \qq &\sin^2\theta_w=3/8,\cr
 m_t=2\ m_W, \qq & m_H=3.14\ m_W. \label{1}\ee
 All four
relations are unstable under quantum corrections
\cite{ksza} and raise the question of how to quantize a
field theory of non-commutative geometry. If
interpreted at their natural scale $m_W$, the first two
relations are in contradiction with experiment, the
third is close to the recently announced  top mass
\cite{top}. When calculating with a more general
scalar product \cite{ks} one still gets one relation
among coupling constants,
\bb
\frac{1}{3}\frac{1}{\alpha_3}+
\frac{.25-\sin^2\theta_w}{\alpha_{em}}=0,\ee
and a conflict with experiment. Connes and
Lott \cite{cl34} also wrote down the most general gauge
invariant scalar product. Due to the
high degree of reducibility of the fermion
representations in the standard model, the general
scalar product destroys all four relations, however
leaving a relation between the top and the Higgs
masses and leaving an inequality for $\sin^2\theta_w$.
There is a natural subclass of scalar products, that
determines the top and Higgs masses as in equations
(\ref{1}). E.
Alvarez,  J. M. Gracia-Bond\'\i a \& C. P. Mart\'\i n
\cite{agm} have carried out a renormalisation group
analysis of these two mass relations in ordinary
quantum field theory. They find a weak scale
dependence only.

The purpose of this article is to give the computational
details of the standard model with general scalar
product and to discuss the phenomenological
implications. The more mathematically inclined
reader is referred to a companion paper \cite{ks3}.

\section{The input of the standard model in the
Connes-Lott scheme}

 The standard model in non-commutative geometry is
described by two real algebras, one for electro-weak
interactions: $\aa:=\hhh\op\cc$ with group of
unitaries $SU(2)\times U(1)$, and one for strong
interactions: $\aa':=M_3(\cc)\op\cc$ with group of
unitaries $U(3)\times U(1)$. We denote by $\hhh$ the
algebra of quaternions. Its elements are complex
$2\times 2$ matrices of the form
\bb \pp{x&-\bar y\cr y&\bar x},\qq x,y\in\cc.\ee
Both algebras $\aa$ and $\aa'$  are represented on
the same Hilbert space $\hh=\hh_L\op\hh_R$ of left
and right handed fermions,
\bb\hh_L= \left(\cc^2\ot\cc^N\ot\cc^3\right)\ \op\
\left(\cc^2\ot\cc^N\right),\ee
\bb\hh_R=\left((\cc\op\cc)\ot\cc^N\ot\cc^3\right)\
\op\ \left(\cc\ot\cc^N\right).\ee
The first factor denotes weak isospin, the second $N$
generations, $N=3$, and the third denotes colour
triplets and singlets. With respect to the standard basis
\bb \pp{u\cr d}_L,\ \pp{c\cr s}_L,\ \pp{t\cr b}_L,\
\pp{\nu_e\cr e}_L,\ \pp{\nu_\mu\cr\mu}_L,\
\pp{\nu_\tau\cr\tau}_L\ee
of $\hh_L$ and
\bb\matrix{u_R,\cr d_R,}\qq \matrix{c_R,\cr s_R,}\qq
\matrix{t_R,\cr b_R,}\qq  e_R,\qq \mu_R,\qq
\tau_R\ee
 of $\hh_R$, the representations are given
by block diagonal matrices. For $(a,b)\in\hhh\op\cc$
we set
\bb B:=\pp{b&0\cr 0&\bar b}\ee
and define a representation of $\aa$ by
\bb\rho(a,b):=\pp{
a\ot 1_N\ot 1_3&0&0&0\cr
0&a\ot 1_N&0&0\cr
0&0&B\ot 1_N\ot 1_3&0\cr
0&0&0&\bar
b1_N}=\pp{\rho_L(a)&0\cr0&\rho_R(b)}\ee
and for
$(c,d)\in M_3(\cc)\op\cc$ we define a $\aa'$
representation
\bb\rho'(c,d):=\pp{
1_2\ot 1_N\ot c&0&0&0\cr
0&d1_2\ot 1_N&0&0\cr
0&0&1_2\ot 1_N\ot c&0\cr
0&0&0&d1_N}.\ee
The last piece of input is the fermion mass matrix
$\mm$ which constitutes the self adjoint `internal
Dirac operator':
\bb\dd&:=&\pp{
0&0&\pp{M_u\ot1_3&0\cr 0&M_d\ot 1_3}&0\cr
0&0&0&\pp{0\cr M_e}\cr
\pp{M_u^*\ot1_3&0\cr 0&M_d^*\ot 1_3}&0&0&0\cr
0&\pp{0&M_e^*}&0&0}\cr\cr \cr
&=:&\pp{0&\mm\cr\mm^*&0}\ee
with
\bb M_u:=\pp{
m_u&0&0\cr
0&m_c&0\cr
0&0&m_t},\qq M_d:= C_{KM}\pp{
m_d&0&0\cr
0&m_s&0\cr
0&0&m_b},\qq M_e:=\pp{
m_e&0&0\cr
0&m_\mu&0\cr
0&0&m_\tau}\ee
where $C_{KM}$ denotes the
Cabbibo-Kobayashi-Maskawa matrix. All indicated
fermion masses are supposed positive and different.
Note that the
 strong interactions are vector-like: the chirality
operator
 \bb \chi=\pp{
-1_2\ot 1_N\ot 1_3&0&0&0\cr
0&-1_2\ot1_N&0&0\cr
0&0&1_2\ot 1_N\ot 1_3&0\cr
0&0&0&1_N}\ee
and the `Dirac operator' commute with $\aa'$
\bb \left[\dd,\rho'(\aa')\right]=0,\ee
\bb \left[\chi,\rho'(\aa')\right]=0.\ee

\section{The Connes-Lott model building kit, internal
space}

With this input --- an involution algebra $\aa$, a
faithful representation $\rho$ of $\aa$ on $\hh$, that
decomposes into a left handed representation $\rho_L$
on $\hh_L$ and a right handed one, and a `Dirac
operator' $\dd$ --- Connes constructs the central piece
of his model building kit, a differential algebra
$\Omega_\dd\aa$. This construction may seem
complicated at first sight, but it has profound roots in
non-commutative geometry.

It starts with an auxiliary differential algebra
$\Omega \aa,$
the so called universal differential envelope of $\aa$:
\bb   \Omega^0\aa := \aa,\ee
$\Omega^1\aa$ is
generated by symbols $\delta a$, $ a \in \aa$ with
relations
\bb   \delta 1 = 0 \ee
\bb    \delta(aa') = (\delta a)a'+a\delta a'.\ee
For the moment $\aa$ is an arbitrary involution
algebra with generic elements $a,\ a',...$ Forget about
quaternions and the second algebra $\aa'$.
 $\Omega^1\aa$ consists
of finite sums of terms of the form $a_0\delta a_1,$
\bb   \Omega^1\aa = \left\{ \sum_j a^j_0\delta
a^j_1,\quad a^j_0, a^j_1\in \aa\right\}\ee
and likewise
for higher $p$
\bb   \Omega^p\aa = \left\{ \sum_j
a^j_0\delta a^j_1...\delta a^j_p ,\quad a^j_q\in
\aa\right\}.\ee
 The differential $\delta$ is defined by
\bb   \delta(a_0\delta a_1...\delta a_p) :=
   \delta a_0\delta a_1...\delta a_p.\ee

The involution $^*$ is
extended from the algebra $\aa$ to
 $\Omega^1\aa$ by putting
\bb   (\delta a)^* := \delta(a^*) =:\delta a^*.\ee
 With the definition
\bb (\varphi\psi)^*=\psi^*\varphi^*\ee
the involution is extended to the whole
differential envelope.

The next step is to extend the
representation $\rho:=\rho_L\op\rho_R$ on
$\hh:=\hh_L\op\hh_R$ from the  algebra $\aa$   to its
universal differential envelope $\Omega\aa$. This
extension deserves a new name:
\bb  \pi :
\Omega\aa  \longrightarrow&{\rm End}(\hh)
   \eee
 \bb\pi(a_0\delta a_1...\delta a_p) :=
(-i)^p\rho(a_0)[\dd,\rho(a_1)] ...[\dd,\rho(a_p)]
\label{pi}.\ee
A straightforward calculation shows that $\pi$ is in
fact a  representation of $\Omega\aa$ as involution
algebra, and we are tempted to define also a
differential, again denoted by $\delta$,
 on $\pi(\Omega\aa)$ by
\bb
\delta\pi(\hat\varphi):=\pi(\delta\hat\varphi).
\label{trial}\ee
However, this definition does not make sense if there
are forms
$\hat\varphi\in\Omega\aa$ with
$\pi(\hat\varphi)=0$ and  $\pi(\delta\hat\varphi)
\not= 0$. By dividing out these unpleasant forms,
we finally arrive at the differential algebra
$\Omega_\dd\aa$, the real thing.
\bb   \Omega_\dd\aa :=
{{\pi\left(\Omega\aa\right)}\over J}\ee
with
\bb   J := \pi\left(\delta\ker\pi\right) =:
\bigoplus_p J^p,\ee
($J$ for junk).
On the quotient now,
the differential (\ref{trial}) is well defined. Degree by
degree we have:  \bb   \Omega_\dd^0\aa =\rho(\aa)\ee
because $J^0=0$ ,
\bb   \Omega_\dd^1\aa = \pi(\Omega^1\aa)\ee
because $\rho$ is faithful,
and in degree $p\geq2$
\bb   \Omega_\dd^p\aa =
{{\pi(\Omega^p\aa)}\over
{\pi(\delta(\ker\pi)^{p-1})}}.\ee
While $\Omega\aa$ has no cohomology,
$\Omega_\dd\aa$  in general does. In fact, in infinite
dimensions,
if $\ff$ is the algebra of complex functions on
spacetime $M$ represented on the square integrable
spinors by multiplication and if $\dd$ is the genuine
Dirac operator then $\Omega_\dd\ff$ is de Rham's
differential algebra of differential forms on $M$.

We come back to our finite
dimensional case. Remember that the elements of the
auxiliary differential algebra $\Omega\aa$ that
we introduced for book keeping purposes only, are
abstract entities defined in terms of symbols and
relations. On the other hand the elements of
$\Omega_\dd\aa$, the ``forms'', are operators on the Hilbert
space $\hh$, i.e. concrete matrices of
complex numbers. Therefore there is a
natural scalar product defined by
\bb   <\hat\varphi,\hat\psi> :=
\t (\hat\varphi^*\hat\psi),
\quad  \hat\varphi, \hat\psi \in
\pi(\Omega^p\aa)\label{sp}\ee
for elements of equal degree and by zero for two
elements of different degree.
With this scalar product  $\Omega_\dd\aa$ is a
subspace of $\pi(\Omega\aa)$, by  definition
orthogonal to the junk. As a subspace
$\Omega_\dd\aa$  inherits a scalar product which deserves
a special name ( , ). It is  given by
\bb   (\varphi,\psi) =
<\hat\varphi,P\hat\psi>, \quad
\varphi, \psi \in \Omega_\dd^p\aa\ee
 where $P$ is the orthogonal
projector in $\pi(\Omega\aa)$  onto the
ortho-complement of $J$ and $\hat\varphi$ and
$\hat\psi$ are any  representatives in their classes.
Again the scalar product vanishes  for forms with
different degree. For real algebras all traces must be
understood as real part of the trace.

Let us remark the
existence of a natural subclass of scalar products
\cite{cl34} defined by elements $z$, that are not only in
the commutant of $\aa$ but are taken from image
under $\rho$ of the center of $\aa$.

Now suppose that the left and right representations
are reducible as the case in the standard model. Then
there is an obvious generalization of the scalar product
(\ref{sp}).
It is  constructed by taking the trace over each
irreducible part of $\hh$ separately and by
multiplying each trace by an independent positive
constant. The most general scalar product in this
context reads \cite{cl34}
\bb   <\hat\varphi,\hat\psi> :=
\t (\hat\varphi^*\hat\psi z),
\quad  \hat\varphi, \hat\psi \in
\pi(\Omega^p\aa)\label{gsp}.\ee
$z$ is any positive operator on $\hh$, that commutes
with $\rho(\aa)$ and with $\dd$.

At this stage, there is a first contact with gauge
theories.  Consider the vector space of anti-Hermitian
1-forms
\bb   \left\{ H\in \Omega_\dd^1\aa,\ H^*=-H
\right\}.\ee
Let us call these elements
Higgses. The space of
Higgses carries an affine  representation of
the group of unitaries
\bb    G = \{g\in \aa,\ gg^\ast
=g^*g=1\}\ee
defined by
\bb H^g &:=&\
\rho(g)H\rho(g^{-1})+\rho(g)\delta (\rho(g^{-1})) \cr
       &=&\ \rho(g)H\rho(g^{-1})+(-i)\rho(g)[\dd,\rho
(g^{-1})] \cr
      &=&\ \rho(g)[H-i\dd]\rho(g^{-1})+i\dd.\ee
 $H^g$ is
the `gauge transformed of $H$'.  This transformation
law makes the Higgs play the role of a (finite
dimensional) gauge potential. In fact every $H$ defines
a covariant derivative  $\delta +H$, covariant under the
left action of $G$ on $\Omega_\dd\aa$:
\bb   ^g\psi := \rho(g)\psi, \quad
\psi\in\Omega_\dd\aa\ee
 which means
\bb   (\delta+H^g)\
^g\psi = \ ^g\lb(\delta+H)\psi\rb.\ee
 Also we define the
curvature $C$ of $H$ by
\bb   C := \delta H+H^2\ \in\Omega_\dd^2\aa.\ee
Note that here and later $H^2$ is considered as element
of $\Omega_\dd^2\aa$ which means it is the projection $P$
applied to $H^2\in \pi(\Omega^2\aa)$.
The curvature $C$ is a Hermitian 2-form with {\it
homogeneous} gauge transformations
 \bb   C^g :=
\delta(H^g)+(H^g)^2 = \rho(g) C \rho(g^{-1}).\ee

Finally we define the preliminary Higgs potential
$V_0(H)$, a functional on the  space of Higgses, by
\bb   V_0(H)
:= (C,C) = \t[(\delta H+H^2)P(\delta H+H^2)].\ee
It is a
polynomial of degree 4 in $H$ with real, non-negative
values.  Furthermore it is gauge invariant,
$V_0(H^g)= V_0(H)$,
 because of the homogeneous transformation
property of the  curvature $C$ and because the
orthogonal projector $P$ commutes  with all gauge
transformations,
$\rho(g)P =P\rho(g)$.
The most remarkable property of the preliminary
Higgs potential is that, in most cases, its minimum
spontaneously breaks the group $G$. To see this, we
introduce the change of variables
\bb\Phi:= H-i\dd.\ee
 This variable transforms homogeneously:
\bb\Phi^g=H^g-i\dd
=\rho(g)[H-i\dd]\rho(g^{-1})+i\dd-i\dd
=\rho(g)\Phi\rho(g^{-1})\label{hom}.\ee
Now $H=0$, or
equivalently $\Phi=-i\dd$, is certainly a
minimum of the preliminary Higgs potential and this
minimum spontaneously breaks $G$ if it is gauge
variant.

The invariance group of the Higgs potential is
the group of unitaries $G$, a subset of the algebra
$\aa$. $G$ can be reduced to a special subgroup
by means of a so called unimodularity condition. These
conditions are defined on $G^0$, the connected
component of the identity in $G$. For a finite
dimensional algebra $\aa$ represented on a finite
dimensional Hilbert space $\hh$, the unimodularity
conditions take a simple form. Every element $g\in
G^0$ can be written
\bb g=e^X,\ee
where $X$ is an element in the Lie algebra $\gg$ of
$G$. The Lie algebra $\gg$ is again a subset of the
algebra $\aa$,
\bb \gg=\left\{X\in\aa,\ X^*+X=0\right\}. \ee
Choose an
element $p$ in the center of $\aa$ such that $\t
\rho(p)\in\zz$, $p$ stands for projection. For every
$p$, there is a unimodularity condition
\bb \t\rho(Xp)=0\ee defining a subgroup of $G^0$,
\bb G_p:=\left\{g=e^X\in G^0,\
\t\rho(Xp)=0\right\}.\ee

\section{The internal space of the standard model}

We now apply the construction outlined above to the
standard model. Obviously, the standard model is not
the right example to get familiar with the Connes-Lott
scheme. Miraculously enough, the standard
model contains the minimax example, analogue of
the Georgi-Glashow $SO(3)$  model \cite{gg} in the
Yang-Mills-Higgs scheme (a maximum of
pleasure with a minimum of effort). This example
represents the electro-weak algebra $\aa=\hhh\op\cc$
on $two$ generations of leptons. Its only drawback are
neutrinos with electric charge, a drawback, that can be
corrected by adding strong interactions.

Anyway, let us start the computation of the
differential algebra $\Omega_\dd\aa$ for the electro-weak
algebra with
generic element $(a,b)\in\hhh\op \cc$ represented
on the long list of fermions. A general
1-form is a sum of terms
\bb
\pi((a_0,b_0)\delta(a_1,b_1)) =-i \pmatrix
{0&\rho_L(a_0)\left(\mm\rho_R(b_1)-\rho_L(a_1)
\mm\right)\cr \rho_R(b_0)\left(\mm^*\rho_L(a_1)-
\rho_R(b_1) \mm^*\right)&0}\ee
 and as
vector space
\bb   \Omega_\dd^1\aa = \left\{i\pmatrix
{0&\rho_L(h)\mm\cr \mm^*\rho_L(\tilde h^*)&0},\
h,\tilde h\in \hhh\right\}.\ee
The Higgs being an anti-Hermitian 1-form
\bb H=  i\pmatrix
{0&\rho_L(h)\mm\cr \mm^*\rho_L(h^*)&0},\qq
h=\pp{h_1&-\bar h_2\cr h_2&\bar h_1}\in \hhh\ee
is parameterized by one complex doublet
\bb \pp{h_1\cr h_2},\qq h_1,h_2\in\cc.\ee
The junk in degree two turns out to be
\bb J^2=\left\{i\pp{j\ot\Delta&0\cr 0&0},\qq
j\in\hhh\right\}\ee
with
\bb\Delta:=\frac{1}{2}\pp{\left(M_uM_u^*-M_dM_d^*
\right)\ot 1_3&0\cr
0&-M_eM_e^*}.\ee
To project it out, we use the general scalar product
(\ref{gsp}) with the real part of the trace.
Here the most general $z$, that commutes with
$\rho(\aa)$ and $\dd$,  has the form
\bb z=\pp{ 1_2\ot 1_N\ot x&0&0&0\cr
0&1_2\ot y&0&0\cr
0&0&1_2\ot 1_N\ot x&0\cr
0&0&0&y} \ee
where $y$ is a positive, diagonal $N\times N$ matrix
and $x$ is a
positive $3\times 3$ matrix. Note that this $z$ also
commutes with the chirality operator $\chi$. The
scalar product defined with this $z$ has a natural
interpretation. Indeed, we shall see later that, without
loss of generality, we may take $x$ to be a positive
multiple of the identity. Then, the general scalar
product is just a sum of the simplest scalar products
in each irreducible part of the fermion
representation, each weighted with a separate positive
constant. We have four irreducible parts, the three
lepton families and all quarks together. Due to the
Cabbibo-Kobayashi-Maskawa mixing, the ponderations
of the three quark families are identical. If, in addition,
we suppose that $z$ lie in $\rho$(center$\aa$) then
we have $x=\lambda 1_3$, $y=\lambda 1_N$ with a
positive constant $\lambda$.

 With respect to the general scalar product, we can
write the 2-forms as
\bb \Omega_\dd^2\aa=\left\{\pp{
\tilde c\ot \Sigma &0\cr
0&\mm^*\rho_L(c)\mm},\qq
\tilde c,c\in\hhh\right\}\ee
with
\bb \Sigma:=\frac{1}{2}\pp{\left(M_uM_u^*+M_dM_d^*
\right)\ot 1_3&0\cr
0&M_eM_e^*}.\ee
Since $\pi$ is a
homomorphism of involution algebras, the product  in
$\Omega_\dd\aa$ is given by matrix multiplication
followed by the orthogonal projection $P$ and
the involution  is given by transposition complex
conjugation. In order to
calculate the  differential $\delta$, we go
back to the universal differential envelope. The result
is
\bb\delta :
\Omega_\dd^1\aa&\longrightarrow&\Omega_\dd^2\aa\cr
\nobreak\cr
i\pmatrix
{0&\rho_L(h)\mm\cr \mm^*\rho_L(\tilde h^*)&0}
 &\longmapsto& \pp{
\tilde c\ot \Sigma&0\cr
0&\mm^*\rho_L(c)\mm}\ee
with
\bb\tilde c=c=h+\tilde h^*.\ee
We are now in position to compute the curvature and
the preliminary Higgs potential:
\bb C:=\delta H+H^2=
\left(1-|\varphi|^2\right)\pp{1\ot\Sigma&0 \cr
0&\mm^*\mm}\ee
where we have introduced the {homogeneous} scalar
variable
\bb \Phi:= H-i\dd=:i\pmatrix
{0&\rho_L(\varphi)\mm\cr
\mm^*\rho_L(\varphi^*)&0},\qq
\varphi=\pp{\varphi_1&-\bar \varphi_2\cr
\varphi_2&\bar \varphi_1}\in \hhh,\ee
 \bb |\varphi|^2:=|\varphi_1|^2+|\varphi_2|^2.\ee
The preliminary Higgs potential
 \bb V_0=\t\left[C^2\right]=
 \left(1-|\varphi|^2\right)^2&\times&\left(
2\t\left[\left(M_u^*M_u\right)^2\right]\t x+
2\t\left[\left(M_d^*M_d\right)^2\right]\t x\right.
\cr &&\qq+
\t\left[M_u^*M_uM_d^*M_d\right]\t x+
\t\left[M_d^*M_dM_u^*M_u\right]\t x\cr&&\left.\qq +
2\t\left[\left(M_e^*M_e\right)^2y\right]\right)\ee
breaks the $SU(2)\times U(1)$ symmetry down to
$U(1)$.

Finally we must compute the differential algebra
$\Omega_\dd\aa'$ of the strong algebra. As the strong
interactions are vector-like this is trivial:
\bb \Omega_\dd^0\aa'&=&\rho'(\aa'),\cr
\Omega_\dd^p\aa'&=&0,\qq p\geq 1.\ee
Consequently there is no Higgs and no Higgs
potential in the strong internal space. For later use, we
still need the general positive operator $z'$ on $\hh$,
that commutes with $\rho'(\aa')$ and with the internal
Dirac operator $\dd$:
\bb z'=\pp{
\pp{r&0\cr 0&C_{KM}\,s\,C_{KM}^*}\ot1_3
&0& \pp{k&0\cr 0&C_{KM}p}\ot 1_3&0\cr
0&\pp{u&0\cr 0&v}&0&\pp{0\cr w}\cr
\pp{k&0\cr 0&p C_{KM}^*}\ot 1_3&0&\pp{r&0\cr
0&s}\ot1_3&0\cr  0&\pp{0&w}&0&v}\ee
where $r$, $s$, $u$, $v$, $k$, $p$ and $w$ are
Hermitian, $N\times N$ matrices. All of them with
exception of $u$ are  diagonal. If in addition
we suppose that $z'$ lie in $\rho'$(center$\aa'$) then
we have $r=s=\lambda_q1_N$, $u=v=\lambda_\ell 1_N$
with  positive constants $\lambda_q$, $\lambda_\ell$
and we have $k=p=w=0$. Note that a
general $z'$ in the commutant does not commute with
the chirality operator unless we set $k=p=w=0$.

Thomas Schucker
CPT, case 907
F-13288 Marseille
cedex 9
tel.: (33) 91 26 95 32
fax:  (33) 91 26 95 53

\section{The Connes-Lott model building kit,
 spacetime added}

 In this section, the Higgses $H$
are promoted to genuine fields, i.e. spacetime
dependent vectors. As already in classical quantum
mechanics, this promotion is achieved by tensorizing
with functions. Let us denote by $\ff$ the algebra of
(smooth, real or complex valued) functions over
spacetime $M$. Consider the algebra $\aa_t:=\ff\ot\aa$.
The group of unitaries of the tensor algebra $\aa_t$ is
the gauged version of the group of unitaries of the
internal algebra $\aa$, i.e. the group of functions
from spacetime into the group $G$. Consider the
representation $\rho_t:=\ul\cdot\ot\rho$ of the
tensor algebra on the tensor product
$\hh_t:=\sss\ot\hh,$
where $\sss$ is the Hilbert space of square integrable
spinors on which functions act by multiplication:
$ (\ul f\psi)(x):=f(x)\psi(x)$, $ f\in\ff,\
\psi\in\sss$. We denote the genuine Dirac operator
by $\ddd$ and its chirality operator by $\gamma_5$.
The definition of the tensor product of Dirac operators,
\bb\dd_t:=\ddd\ot 1+\gamma_5\ot\dd\ee
comes from non-commutative geometry. We now
repeat the above construction for the infinite
dimensional algebra $\aa_t$ with representation
$\rho_t$ and Dirac operator $\dd_t$. As
already stated, for $\aa=\cc,\ \hh=\cc,\ \mm=0$, the
differential algebra $\Omega_{\dd_t}\aa_t$ is
isomorphic to the de Rham algebra of differential
forms $\Omega (M,\cc)$. For general $\aa$, using the
notations of  \cite{sz}, an anti-Hermitian 1-form
\bb H_t\in\Omega_{\dd_t}^1\aa_t,\quad
H^*_t=-H_t\eee  contains two pieces, an anti-Hermitian
Higgs {\it field} $
H\in\Omega^0(M,\Omega_\dd^1\aa)$ and a genuine
gauge field $ A\in\Omega^1(M,\rho(\gg))$
with values in the Lie algebra of the group of
unitaries,
 \bb \gg:=\left\{ X\in\aa,\ X^*+X=0\right\},\ee
represented on $\hh$. The curvature of $H_t$
\bb C_t:=\delta_tH_t+H_t^2\
\in\Omega_{\dd_t}^2\aa_t\ee
contains three pieces,
\bb C_t=C+F-\dee\Phi\gamma_5,\ee
 the ordinary, now $x$-dependent
curvature $C=\delta H+H^2$, the field strength
\bb F:=\de A+\frac{1}{2}[A,A]\quad \in
\Omega^2(M,\rho(\gg))\ee
and the covariant derivative of $\Phi$
\bb\dee \Phi=\de \Phi+[A\Phi-\Phi A]
\quad\in\Omega^1(M,\Omega_\dd^1\aa).\ee
Note that the covariant derivative may be applied to
$\Phi$ thanks to its homogeneous transformation law,
equation (\ref{hom}).

The definition of the Higgs potential in the
infinite dimensional space
\bb V_t(H_t):=(C_t,C_t)\ee
requires a suitable regularisation of the sum of
eigenvalues over the space of spinors $\sss$.
Here we have to suppose spacetime to be compact and
Euclidean. Then, the regularisation is
achieved by the Dixmier trace which allows an explicit
computation of $V_t$.
One of the miracles in the Connes-Lott scheme is that
$V_t$ alone reproduces the complete bosonic action of
a Yang-Mills-Higgs model. Indeed, it consists of three
pieces,  the Yang-Mills action, the covariant
Klein-Gordon action and an integrated Higgs potential
\bb V_t(A+H)=\int_M\t \left(F*F\,z\right)+ \int_M\t
\left(\dee\Phi^**\dee\Phi\, z\right)+
\int_M*V(H).\label{Vt}\ee
As the preliminary Higgs potential $V_0$, the (final)
Higgs potential $V$ is calculated as a function of the
fermion masses,
   \bb V:=V_0-\t[\alpha C^*\alpha C \,z]=
\t[(C-\alpha C)^*(C-\alpha C)\,z],\ee
where the linear map
\bb   \alpha: \Omega_\dd^2\aa\longrightarrow
         \rho(\aa)+\pi(\delta(\ker\pi)^1)\ee
is determined by the two equations
\bb
\t\lb R^*(C-\alpha C)\,z\rb&=&0\qquad{\rm for\ all}\
R\in\rho(\aa), \label{a1}\\
 \t\lb K^*\alpha C\,z\rb &=&0\qquad {\rm for\ all}\
K\in\pi(\delta(\ker\pi)^1).\label{a2}\ee
All remaining traces are over the finite dimensional
Hilbert space $\hh$. We denote the Hodge star
by $*\cdot$. It should not be confused with the
involution $\cdot^*$. Note the `wrong' relative sign of
the third term in equation (\ref{Vt}). The sign is in fact
correct for an Euclidean spacetime.

A similar miracle happens in the fermionic sector,
where the completely covariant action
 $\psi^*(\dd_t+iH_t)\psi$
reproduces the complete fermionic action of a
Yang-Mills-Higgs model.
We denote by
 \bb\psi=\psi_L+\psi_R\ \in
\hh_t=\sss\ot\,\left(\hh_L\op\hh_R\right)\eee
 the multiplets of spinors and by $\psi^*$ the dual of
$\psi$ with respect to the scalar product of the
concerned Hilbert space. For the purpose of this
general section, we set
\bb H=:i\pmatrix {0&\tilde h \cr
\tilde h^*&0}\in\Omega_\dd^1\aa,\ee
\bb\Phi=  H-i\dd=:i\pmatrix {0&\tilde\varphi \cr
\tilde\varphi^*&0}\in\Omega_\dd^1\aa.\ee
Then
\bb\psi^*(\dd_t+iH_t)\psi&=&
\int_M*\psi^*(\ddd+i\gamma(A))\psi
-\int_M*\left(\psi_L^*\tilde h\gamma_5\psi_R
+\psi_R^*\tilde h^*\gamma_5\psi_L\right)\cr\cr
&&+\int_M*\left(\psi_L^*\mm\gamma_5\psi_R
+\psi_R^*\mm^*\gamma_5\psi_L\right)\cr\cr
&=&\int_M*\psi^*(\ddd+i\gamma(A))\psi
-\int_M*\left(\psi_L^*\tilde \varphi\gamma_5\psi_R
+\psi_R^*\tilde
\varphi^*\gamma_5\psi_L\right)\label{diract}\ee
containing the ordinary Dirac action
and the Yukawa couplings. Note the
unusual appearance of $\gamma_5$ in the fermionic
action (\ref{diract}). Just as the `wrong' signs in the
bosonic action (\ref{Vt}), these $\gamma_5$ are
proper to the Euclidean signature and disappear in the
Minkowski signature.

We close this section with a word of caution. In fact, we
have slightly over-simplified the outline of the
Connes-Lott scheme. The omitted details can be found
in reference \cite{is2}. They are irrelevant for our
present purpose, the standard model to which we return
now.

\section{The standard model, spacetime added}

Let us apply the construction outlined above to the
standard model. Recall the expression of the
curvature in the electro-weak sector
\bb C=
\left(1-|\varphi|^2\right)\pp{1_2\ot\Sigma&0 \cr
0&\mm^*\mm}.\ee
A straightforward application of equations (\ref{a1},
\ref{a2}) --- taking the real part of the traces is
understood --- yields the projection $\alpha C$. It is
again block diagonal with diagonal elements:
\bb \alpha C_{qL}&=&
\frac{1-|\varphi|^2}{2}\frac{
\t\left[M_u^*M_u\right]\t x+
\t\left[M_d^*M_d\right]\t x+
\t\left[(M_e^*M_e\,y\right]}{N\t x+\t y}\,
1_2\ot 1_N\ot 1_3\\ \cr
\alpha C_{\ell L}&=&
\frac{1-|\varphi|^2}{2}\frac{
\t\left[M_u^*M_u\right]\t x+
\t\left[M_d^*M_d\right]\t x+
\t\left[(M_e^*M_e\,y\right]}{N\t x+\t y}\,
 1_2\ot 1_N\\ \cr
\alpha C_{qR}&=&
\frac{1-|\varphi|^2}{2}\frac{
\t\left[ M_u^*M_u\right]\t x+
\t\left[M_d^*M_d\right]\t x+
\t\left[(M_e^*M_e\,y\right]}{N\t x+\t y/2}\,
 1_2\ot 1_N\ot 1_3\\ \cr
\alpha C_{\ell R}&=&
\frac{1-|\varphi|^2}{2}\frac{
\t\left[M_u^*M_u\right]\t x+
\t\left[M_d^*M_d\right]\t x+
\t\left[(M_e^*M_e\,y\right]}{N\t x+\t y/2}\, 1_N.\ee
The Higgs potential is computed next,
\bb V= K\left(1-|\varphi|^2\right)^2,\ee
\bb K&:=&\frac{3}{2}
\t\left[\left(M_u^*M_u\right)^2\right]\t x+
\frac{3}{2}
\t\left[\left(M_d^*M_d\right)^2\right]\t x
\cr &&+
\t\left[M_u^*M_uM_d^*M_d\right]\t x
\cr &&
+\frac{3}{2}
\t\left[M_e^*M_eM_e^*M_e\,y\right]\cr &&
-\frac{1}{2} L^2\left[\frac{1}{N\t x+\t y}+
\frac{1}{N\t x+\t y/2}\right],\ee

\bb L:=\t\left[M_u^*M_u\right]\t x+
\t\left[M_d^*M_d\right]\t x+
\t\left[(M_e^*M_e\,y\right].\ee
Note that the scalar fields $\varphi_1$ and $\varphi_2$
are not properly normalized, they are dimensionless.
To get their normalization straight we have to compute
the factor in front of the kinetic term in the
Klein-Gordon action:
\bb \t \left(\de\Phi^**\de\Phi\, z\right)=*2L|\partial
\varphi|^2.\ee
Likewise, we need the normalization of the
electro-weak gauge bosons:
\bb \t \left(F*F\,z\right)=*\left(N\t x+\t y\right)
\left(\partial_\mu W_\nu^+\,\partial^\mu
W^{-\nu}-...\right). \ee
We end up with the following mass relations:
\bb m_W^2&=&\frac{L}{N\t x+\t y} \label{w}\\ \cr
m_H^2&=&\frac{2K}{L}.\label{h}\ee

Finally, we turn to the relations among coupling
constants. They are due to the fact that the gauge
invariant scalar product on the internal Lie algebra,
the Lie algebra of the group of unitaries
$ \gg:=\left\{ X\in\aa,\ X^*+X=0\right\}$,
in the Yang-Mills action (\ref{Vt}) is not general
but stems from the trace over the fermion
representation $\rho$ on $\hh$. Since this
representation is faithful the scalar product (\ref{sp})
indeed induces an invariant scalar product on $\gg$.

The fact that the standard model can be written in the
setting of non-commutative geometry depends
crucially, at this point, on two happy circumstances.
Firstly, the electric charge `generator'
\bb Q=\pp{
\pp{2/3&0\cr 0&-1/3}\ot 1_N\ot 1_3&0&0&0\cr
0&\pp{0&0\cr 0&-1}\ot 1_N&0&0\cr
0&0&\pp{2/3&0\cr 0&-1/3}\ot 1_N\ot 1_3&0\cr
0&0&0&-1_N}\ee
is an element of $i\rho(\gg)\op i\rho'(\gg')$. Indeed it
 is a linear
combination of weak isospin $I_3$ and elements of the
three $u(1)$ factors:
\bb Q=\rho\left(\pp{1/2&0\cr 0&-1/2},0\right)+
\frac{1}{2i}\rho(0,i)+\frac{1}{6i}\rho'(i1_3,0)
-\frac{1}{2i}\rho'(0,i).\ee
We have put `generator' in quotation marks because
 $iQ$ is a Lie algebra element, not $Q$.
The weak angle $\theta_w$ measures the proportion
of weak isospin in the electric charge:
\bb \frac{Q}{|Q|}= \sin\theta_w\,\frac{I_3}{|I_3|}
+\cos\theta_w\,\frac{Y}{|Y|}.\label{sin}\ee
The hypercharge $Y$ is a linear combination of the
three $u(1)$ factors
\bb Y:=\frac{1}{2i}\rho(0,i)+\frac{1}{6i}\rho'(i1_3,0)
-\frac{1}{2i}\rho'(0,i).\ee
Here comes the second happy circumstance,
this particular combination $Y$ is singled out by two
unimodularity conditions. They reduce the
group of unitaries $SU(2)\times U(1)\times U(3)\times
U(1)$ to $SU(2)\times U(1)\times SU(3)$ with the
surviving $U(1)$ generated by the hyper charge.
Indeed, the center of $\aa\op\aa'$ is four dimensional
with basis $p_1,...,p_4$. $p_1:=\rho(1_2,0)$
projects on $\hhh$, $p_2:=\rho(0,1)$ on $\cc$,
$p_3:=\rho'(1_3,0)$ on $M_3(\cc)$, and
$p_4=\rho'(0,1)$ on $\cc'$, and the group of the
standard model is $G_{p_1}\cap G_{p_2}$.

Let us come back to the calculation of the weak angle.
 Equation (\ref{sin}) is a matrix of equations. Let
us take the difference of the two diagonal elements
corresponding to the left handed neutrino and electron:
\bb \frac{1}{|Q|}=\sin\theta_w\,\frac{1}{|I_3|},\ee
\bb \sin^2\theta_w=\frac{(I_3,I_3)}{(Q,Q)}.\ee
The numerator is readily computed,
\bb (I_3,I_3)=
\t\left[\rho\left(\pp{1/2&0\cr 0&-1/2},0\right)^2
\,z\right]=\frac{1}{2}(N\t x+\t y).\ee
We compute the denominator with Pythagoras' kind
help,
\bb (Q,Q)&=&\t\left[\rho\left(\pp{1/2&0\cr 0&-1/2}
,0\right)^2\,z\right]+
\frac{1}{4}\t\left[\rho(0,1)^2\,z\right]\cr
&&+\frac{1}{36}\t\left[\rho'(1_3,0)^2\,z'\right]
+\frac{1}{4}\t\left[\rho'(0,1)^2\,z'\right]\cr \cr
&=&(N\t x+\frac{3}{4}\t y)
+\frac{1}{6}(\t r+\t s)+\frac{1}{2}(\t u/2+\t v).\ee
Finally the mixing angle is given by
\bb \sin^2\theta_w=\frac{N\t x+\t y}
{2N\t x+\frac{3}{2}\t y
+\frac{1}{3}(\t r+\t s)+\frac{1}{2}\t u+\t v}
.\label{s}\ee
In a similar fashion, the ratio
between strong and weak coupling is computed,
\bb \left(\frac{g_3}{g_2}\right)^2=
\frac{(I_3,I_3)}{(C,C)}=\frac{1}{2}\frac{N\t x+\t y}
{\t r+\t s} \label{g}\ee
where
\bb C:=\rho'\left(\pp{1/2&0&0\cr 0&-1/2&0\cr
0&0&0},0\right).\ee
Here $C$ stands for colour not for curvature.

In this calculation $z$ and $z'$ are different in
general, implying that the electro-weak sector
$\rho(\aa)$ is orthogonal to the strong sector
$\rho'(\aa')$. In the special case where $z=z'$ a
different choice is possible:
\bb (a,a'):=\t \left[\rho(a)^*\rho'(a')\,z\right],\qq
a\in\aa,\ a'\in\aa'.\ee
Then the two $U(1)$ factors $\rho(0,1)$ and
$\rho'(0,1)$ are not orthogonal anymore and the value
of $\sin^2\theta_w$ comes out smaller \cite{ks}. This
choice is closer to grand unified models and yields
$\sin^2\theta_w=3/8=.375$ for $z=z'=1$ to be compared
to $\sin^2\theta_w=12/29=.414$ from equation
(\ref{s}).

\section{Conclusions}

Writing the standard model in terms
of non-commutative geometry yields the four
constraints (\ref{w},\ref{h},\ref{s},\ref{g}) for the
$W$ and Higgs masses, the weak mixing angle and the
ratio of strong and weak coupling constants.
Note that the Cabbibo-Kobayashi-Maskawa matrix has
dropped out from the constraints as well as the off
diagonal, chirality mixing terms $k$, $p$, and $w$ in
$z'$.
 Due to the
highly reducible form of the standard model, these four
constraints involve, in the most general case, five
arbitrary, positive parameters,  the three
eigenvalues $y_1,\ y_2,\ y_3$ of the diagonal
matrix $3y/\t x$,
\bb \alpha:=\frac{\t r+\t s}{\t x},\ee
 and
\bb \beta:=\frac{\t u/2+\t v}{\t x}.\ee
With these parameters the first constraint reads
\bb m^2_t&=&3m^2_W-
\left(m^2_b+m^2_c+m^2_s+m^2_d+m^2_u\right)\cr
&&\qq +\frac{y_1}{3}\left(m^2_W-m^2_e\right)
+\frac{y_2}{3}\left(m^2_W-m^2_\mu\right)
+\frac{y_3}{3}\left(m^2_W-m^2_\tau\right)\cr
&\approx&\left(3+\frac{y_1+y_2+y_3}{3}\right)\,
m_W^2. \ee
This approximation is as good as the present day
experimental accuracy in the measurement of the
$W$-mass,
\bb m_W=80.20\pm .26\  GeV,\qq
\sqrt{m_W^2-1/3\,m_b^2}=80.16\ GeV.\ee
For all practical purpose, we therefore have the
inequality
\bb m_t\,>\,\sqrt{3}\, m_W\,>\,\sqrt{3}\, m_e.\ee
Similarly we get from the second constraint
\bb \sqrt{7/3}\,<\,\frac{m_H}{m_t}\,<\,
\sqrt{3}.\ee
Both constraints together determine the Higgs mass as
a function of the top mass:
 \bb m_H\approx\sqrt{11+3R-\frac{8+2R}{7+R}}\ m_W,
\ee
with
\bb R:=\frac{m_t^2-4m_W^2}{m_W^2}\,>\,-1.\ee
$R$ vanishes in the subclass of scalar products coming
from the center. Note that the Higgs mass is an
increasing function of the top mass, while the
renormalisation group analysis yields a slowly
decreasing function \cite{agm}.

The third constraint,
\bb \sin^2\theta_w=\frac{3+\frac{1}{3}\sum y_j}
{6+\frac{1}{2}\sum y_j+\frac{1}{3}\alpha+\beta},\ee
yields an inequality,
\bb
\sin^2\theta_w\,<\,\frac{2}{3}\,\frac{4+R}{5+R}.\ee
The last constraint,
\bb\left(\frac{g_3}{g_2}\right)^2=\frac{3+\frac{1}{3}
\sum y_j}{2\alpha},\ee is empty.

If we take the natural subclass of scalar
products with $z$ and $z'$ in the centers, the
constraints are more stringent. Indeed, we are now
left with only two positive parameters $q$
and $\ell$:
\bb y_1=y_2=y_3=\frac{\lambda}{\lambda\t
1_3/3}=1,\ee
\bb \alpha=\frac{2\lambda_q\t
1_N}{\lambda\t 1_3}=
2\frac{\lambda_q}{\lambda}=:2q,\ee \bb
\beta=\frac{\frac{3}{2}\lambda_\ell\t 1_N}{\lambda\t
1_3}= \frac{3}{2}\frac{\lambda_\ell}{\lambda}
=:\frac{3}{2}\ell,\ee and the constraints read
\bb m_t&=&2\,m_W,\qq R=0,\\
m_H&=&3.14\,m_W,\ee
\bb \sin^2\theta_w=\frac{4}
{\frac{15}{2}+\frac{2}{3}q+\frac{3}{2}\ell}
\,<\,\frac{8}{15}=.533,\ee
\bb\left(\frac{g_3}{g_2}\right)^2=\frac{1}{q}.\ee

The simplest scalar product is obtained from
$z=\rho(1_2,1)$ and $z'=\rho'(1_3,1)$. Then $q=\ell$
and we get \bb \sin^2\theta_w=\frac{12}{29}=.414,\qq
\left(\frac{g_3}{g_2}\right)^2=1.\ee

We should point out that Connes and Lott's most clear
cut 'prediction'  concerns the mass ratio of the $W$
and the $Z$, a unit $\rho$-parameter without any
appeal. But even without such numerical tests, it seems
clear to us that non-commutative geometry is intrinsic
to the standard model. One may very
well formulate and test general relativity using
flat geometry exclusively. Still, we all agree that
Riemannian geometry is the natural setting --- for at
least two reasons independent of personal taste. We
appreciate the use of the powerful computational tools,
that the mathematicians have developed in
Riemannian geometry. Secondly, there are infinitely
more  gravitational theories within Euclidean
geometry. Likewise, commutative geometry is perfectly
sufficient to write down the standard model and to
compute cross sections, still non-commutative geometry
 is superior \cite{is2}. May be one day, we will know the
masses of the top and the Higgs.  And may be then, the
elements $z$ and $z'$, that do not come from the
centers,  will acquire  the status of the cosmological
constant.

 \end{document}